\documentclass{elsart}
\usepackage{mathrsfs}
\usepackage{bbm}
\usepackage{amsmath,amssymb}
\usepackage{theorem}
\usepackage{url}
\usepackage{version}
\usepackage{alltt}
\usepackage{graphicx}
\usepackage{alltt, amssymb,comment}
\usepackage[plainpages=false,pdfpagelabels,
colorlinks=true,citecolor=blue,hypertexnames=false]{hyperref}

\def\N {\mathbb{N}}

\def\K {\ensuremath{\mathbb{K}}}

\def\DFT {\ensuremath{\text{{DFT}}}}
\def\TFT {\ensuremath{\text{{TFT}}}}
\def\hoeven#1{{#1}}
\def\gathen#1{{#1}}

\def\E {\ensuremath{\mathsf{E}}}

\newtheorem{Theorem}{Theorem}

\begin{document}

\begin{frontmatter}
\title{A simple and fast algorithm for computing exponentials of power series}
\author{Alin Bostan}
\ead{Alin.Bostan@inria.fr} 
\address{Algorithms Project, INRIA Paris-Rocquencourt \\ 78153 Le Chesnay Cedex France} \medskip
and
\author{{\'E}ric Schost} 
\ead{eschost@uwo.ca}
\address{ORCCA and Computer Science Department, Middlesex College, \\
University of Western Ontario, London, Canada}

\begin{abstract} 
  As was initially shown by Brent, exponentials of truncated power
  series can be computed using a constant number of polynomial
  multiplications. This note gives a relatively simple algorithm with
  a low constant factor.
\end{abstract}

\begin{keyword}
Algorithms, exponential, power series, fast Fourier transform, Newton iteration.
\end{keyword}

\maketitle 
\date{3 March 2009}
\end{frontmatter}


Let $\K$ be a ring of characteristic zero and let $h$ be in $\K[[x]]$
with $h(0)=0$. The {\em exponential} of $h$ is the power series
$$\exp(h)=\sum_{i \ge 0} \frac{h^i}{i!}.$$ Computing exponentials is
useful for many purposes, such as solving differential
equations~\cite{BrKu78} or recovering a polynomial from the power sums
of its roots~\cite{Schoenhage82}. 

Using Newton iteration, it has been known since Brent's
work~\cite{Brent76} that exponentials could be computed for the cost
of polynomial multiplication, up to a constant factor.  Following this
original result, a series of works aimed at lowering the multiplicative
factor; they all rely on some form of Newton iteration, either of
order 2 (the ``usual'' form of iteration) or of higher order. Remark
that the question of improving constant factors can be asked with
other applications of Newton iteration (power series inversion, square
root,~\dots)~\cite{Schoenhage00,Bernstein04,HaZi04}, but we do not discuss
those here.

As is customary, we assume that the base ring $\K$ supports the Fast
Fourier Transform (as an aside, note that in the Karatsuba
multiplication model, exponential computation has an asymptotic cost
equivalent to that of multiplication~\cite[\S~4.2.2]{Hoeven02}). If
$m\in\N$ is any power of $2$, we suppose that $\K$ contains a $m$th
primitive root of unity $\omega_m$ such that in addition
$\omega_m=\omega_{2m}^2$; also, 2 is a unit in $\K$. We denote by
$\E(m)$ an upper bound on the cost of evaluating a polynomial of
degree less than $m$ at the points
$(1,\omega_m,\dots,\omega_m^{m-1}).$ Using Fast Fourier Transform, we
have $\E(m) \in O(m\log m)$; we also ask that $\E$ satisfies the
super-linearity property $\E(2m) \ge 2\E(m)$.

\begin{Theorem} \label{theo:main} Let $h \in \K[[x]]$, with $h(0)=0$
  and let $n\in \N$ be a power of $2$. Then, starting from $\omega_n$
  and from the first $n$ coefficients of $h$, one can compute the
  first $n$ coefficients of $\exp(h)$ using $16\frac12\E(n)+24 \frac
  14n$ operations in $\K$.
\end{Theorem}
Using Fast Fourier Transform, polynomials of degree less than $n$ can
be multiplied in $3\E(2n)+O(n)$ operations. Hence, we say that an
exponential can be computed for (essentially) the cost of $2\frac 34$
multiplications.  References to previous work given below use the same ratio
``cost of exponential vs. cost of multiplication''.

As documented by Bernstein~\cite{BernsteinWeb}, the initial algorithm
by Brent had cost $7\frac 13$ times that of multiplication. Bernstein
successively reduced the constant factor to $3\frac 49$ and $2\frac
56$~\cite{Bernstein04} using high-order iterations. Recently, van der
Hoeven~\cite{Hoeven06} obtained an even better constant of $2\frac
13$. However, that algorithm (using a high-order iteration) is quite
complex (to wit, the second-order term in the cost estimate is
likely not linear in $n$); we are not aware of an existing
implementation of it.

As to order-2 iterations, Bernstein~\cite{Bernstein04} obtained a
constant of $3\frac 13$, which was superseded by Hanrot and
Zimmermann's $3\frac14$ result~\cite{HaZi04}. The merits of our
algorithm is thus to be a simple yet faster second order iteration. Compared
to van der Hoeven's result, we are asymptotically slower, but we could
expect to be better for a significant range of $n$, due to the
simplicity of our algorithm.

\begin{center}
\begin{figure}[!!!h]
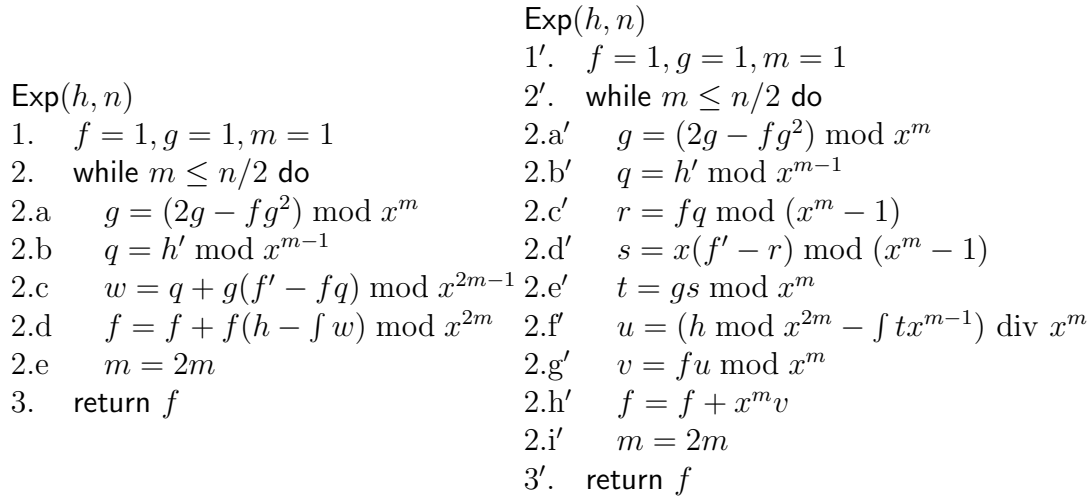

\begin{minipage}{5 cm}
\begin{tabbing}
\quad\quad \= \quad \= \quad \= \quad \kill
{\sf Exp}$(h,n)$\\
1. \> $f=1,g=1,m=1$\\
2. \> {\sf while} $m \leq n/2$ {\sf do}\\
2.a \> \> $g=(2g-fg^2) \bmod x^m$\\
2.b \> \> $q=h' \bmod x^{m-1}$\\
2.c \> \> $w=q+g(f'-fq) \bmod x^{2m-1}$\\
2.d \> \> $f=f+f(h-\int w) \bmod x^{2m}$\\
2.e \> \> $m=2m$\\
3. \> {\sf return} $f$
\end{tabbing}
\end{minipage}
\begin{minipage}{5 cm}
\begin{tabbing}
\quad\quad \= \quad \= \quad \= \quad \kill
{\sf Exp}$(h,n)$\\
1$'$. \> $f=1,g=1,m=1$\\
2$'$. \> {\sf while} $m \leq n/2$ {\sf do}\\
2.a$'$ \> \> $g=(2g-fg^2) \bmod x^m$\\
2.b$'$ \> \> $q=h' \bmod x^{m-1}$\\
2.c$'$ \> \> $r=fq \bmod (x^m-1)$\\
2.d$'$ \> \> $s=x(f'-r) \bmod (x^m-1)$\\
2.e$'$ \> \> $t=gs \bmod x^m$\\
2.f$'$ \> \> $u=(h \bmod x^{2m}-\int t x^{m-1}) {\rm~div~} x^m$\\
2.g$'$ \> \> $v=fu \bmod x^m$\\
2.h$'$ \> \> $f=f+x^m v$\\
2.i$'$ \> \> $m=2m$\\
3$'$. \> {\sf return} $f$
\end{tabbing}
\end{minipage}
\caption{Two versions of the exponential computation}
\label{fig1}
\end{figure}
\end{center}

\paragraph*{Proof.}
For $a=\sum_{i \ge 0} a_i x^i \in \K[[x]]$, we write $a \bmod
x^\ell=\sum_{i=0}^{\ell-1} a_i x^i$ and $a {\rm~div~}
x^\ell=\sum_{i\ge 0} a_{i+\ell} x^i$; computing these quantities does
not require any arithmetic operation. In Figure~\ref{fig1}, we first
give the standard iteration (left), taken from Hanrot and Zimmermann's
note~\cite{HaZi04}, followed by an expanded version where polynomial
multiplications are isolated (right).  Correctness of the left-hand
version is proved in~\cite{HaZi04}; in particular, each time we enter
the loop at Step~2, $f=\exp(h) \bmod x^m$ and $g=1/f \bmod x^{m/2}$ hold.

To prove the correctness of our version, it is enough to show that it
computes the same output as the original one. When entering Step 2 we
have $f = \exp(h) \bmod x^m$; it follows that $x(f'-qf)=0 \bmod x^m$,
with $q=h' \bmod x^{m-1}$.  Since $x(f'-qf)$ has degree less than
$2m$, we deduce that the quantity $s$ of Step 2.d$'$ satisfies
$x(f'-qf)=x^m s$. This implies that $t=gs \bmod x^{m}$ satisfies
$tx^{m-1} = g(f'-q f) \bmod x^{2m-1}$, so that the quantities $w$ of
Step 2.c and $u$ of Step 2.f$'$ satisfy $u=((h -\int w) \bmod
x^{2m}){\rm~div~} x^m$.  The original iteration satisfies $h-\int w =
0 \bmod x^m$, so that actually $x^m u=(h -\int w) \bmod x^{2m}$ and
thus $x^m v=f(h-\int w) \bmod x^{2m}$, with $v=fu \bmod x^m$. The
correctness claim follows.

\medskip
\noindent For $f$ in $\K[x]$ and $m$ a power of 2, we define
$$\DFT(f,m) = (f(1),\dots,f(\omega_m^{m-1})), \quad
\DFT'(f,m) = (f(\omega_{2m}),\dots,f(\omega_{2m}\omega_m^{m-1})),$$ so
that $\DFT(f,2m)$ is, up to reordering, the concatenation of
$\DFT(f,m)$ and $\DFT'(f,m)$. Recall that if $f$ has degree less than
$m$, then $\DFT(f,m)$ can be computed in time $\E(m)$; besides,
$\DFT'(f,m)$ can be computed in time $\E(m)+2m$ (due to the scaling by
$\omega_{2m}$); the inverse DFT in length $m$ can be performed in time
$\E(m)+m$ (due to $m$ divisions by~$m$).

With this, we finally analyze the cost of the algorithm step by
step. We assume that the $n$ elements
$(1,\omega_n,\dots,\omega_n^{n-1})$ have been precomputed in time $n$
once and for all, and stored, so that they are freely available during
the remaining computations.  The hypothesis $\omega_m=\omega_{2m}^2$
ensures that all the needed DFT's solely use (part of) these $n$
elements. 

In what follows, we assume $m$ is a power of $2$, with $m \ge 2$, so
that $m/2$ is an integer.  Recall that at the input of Step 2, $f$ has
degree at most $m-1$ and $g$ has degree at most $m/2 -1$;
additionally, we suppose that $\DFT(g,m)$ is known. Then, the key
ingredients are as follows:
\begin{enumerate}
\item We will compute $\DFT(g,2m)$; since $\DFT(g,m)$ is already
  known, it is enough to compute $\DFT'(g,m)$, which saves a factor of 2.
\item Since $x(f'-qf)=x^m s$, we can compute it modulo $x^m-1$.
\end{enumerate}

\begin{description}
\item [Step 2.a$'$] This step updates $g$ to $1/f \bmod x^m$. The
  product $fg^2$ has degree less than $2m$; it is computed by FFT
  multiplication in length~$2m$. Since $\DFT(g,m)$ is known, we do not
  need to compute $\DFT(g,2m)$ but only $\DFT'(g,m)$.  Hence, the cost
  is \sloppy $\E(2m) \text{\ (DFT of $f$)} + \E(m) + m \text{\ (DFT$'$ of
    $g$)} + 4m \text{\ (pairwise products)} + \E(2m)+ 2m \text{\
    (inverse DFT)} $.

  By the fundamental property of Newton iteration, the first $ m/2 -1$
  coefficients of $g$ and $2g-fg^2$ coincide.  Hence, to deduce
  $2g-fg^2 \bmod x^m$, only $m/2$ sign changes are needed.

\item [Step 2.b$'$] Differentiation takes time $m$; since 
  half of the coefficients were computed at the previous loop,
  the cost can be reduced to $m/2$.

\item [Step 2.c$'$] We compute $r$ by FFT multiplication in length
 $m$. Since $\DFT(f,2m)$, and thus $\DFT(f,m)$, is known, the cost is
 $2\E(m)+2m$.

\item [Step 2.d$'$] Computing $f'-r$ takes time $2m$; 
  multiplication by $x$ modulo $x^m-1$ is free.

\item [Step 2.e$'$] The product $gs$ has degree less than $2m$; it is
  computed by FFT multiplication in length $2m$, of cost
  $3\E(2m)+4m$. This provides $\DFT(g,2m)$, which will be used as
  input in the next iteration.

\item [Step 2.f$'$] Integration and subtraction together take time $2m$.

\item [Step 2.g$'$] The product $fu$ has degree less than $2m$; it is
  computed by FFT multiplication in length $2m$. Since $\DFT(f,2m)$ is
  known, the cost is $2\E(2m)+4m$.

\item [Step 2.h$'$] This step is free.
\end{description}
Hence, the cost of one pass through the main loop is at most
$3\E(m)+7\E(2m) + 22m$. At the last iteration, with $m=n/2$, savings
are possible at Step 2.e$'$, since we do not need to precompute
$\DFT(g,2m)$ for the next iteration. To compute $t = gs\bmod x^m$, we
write
$$g=g_0 +x^{m/2} g_1, \quad s=s_0 +x^{m/2} s_1,\quad
t = g_0 s_0 + x^{m/2} (g_0 s_1 + g_1 s_0) \bmod x^m.$$ We compute $g_0
s_0$ and $g_0 s_1 + g_1 s_0$ by FFT's of order $m$.  Since
$\DFT(g_0,m)$ is known, we just need to compute $\DFT(g_1,m)$,
$\DFT(s_0,m)$ and $\DFT(s_1,m)$, as well as 2 inverse DFT's, for a
cost of $5\E(m)+2m$; the other linear costs (inner products and
additions) sum up to $4\frac12m$.  Adding all costs gives the claimed
complexity result in Theorem~\ref{theo:main}.

\paragraph*{The case of arbitrary $n$.}
We gave our algorithm for $n$ a power of $2$ (the algorithm
of~\cite{HaZi04} does not have this restriction, but assumes that
Fourier transforms can be performed at arbitrary lengths $n$).  We
describe here possible workarounds for the general case.

For an arbitrary value of $n$, Newton iteration will compute the approximations
$\exp(h) \bmod x^{m_i}$, where the sequence $(m_i)_{i\ge 0}$ is
defined by $r=\lceil \log_2(n)\rceil$ and $m_i=\lceil n/2^{r-i}
\rceil$, as in~\cite[Ex. 9.6]{GaGe99}, so that $m_i$ is either
$2m_{i-1}$ or $2m_{i-1}-1$ and thus $m_{i-1}=\lceil m_i/2\rceil$.
Then, the algorithm enters Step~2 knowing $f =\exp(h)\bmod x^{m_i}$
and $g=1/f \bmod x^{m_{i-1}}$; it exits Step~2 with $f =\exp(h)\bmod
x^{m_{i+1}}$ and $g=1/f \bmod x^{m_i}$. Depending on the Fourier
Transform model we use, our improvements can be carried over to this
case as well.

In a model which allows Fourier transforms at roots of unity of any
order, our algorithm extends in a rather straightforward manner. As
before, we also suppose that $\DFT(g,m_i)$ is known at the beginning
of Step~2, where now $\DFT$ can be taken at arbitrary order.  Now, the
multiplications at Steps~2.a$'$,~2.c$'$ and~2.g$'$ are done with
transforms of order respectively $2m_i$, $m_i$ and $2m_i$, but that of
Step~2.c$'$ has order $m_{i+1}$ to enable the next iteration.  This
gives $\exp(h) \bmod x^{2m_i}$, and thus $\exp(h) \bmod x^{m_{i+1}}$,
by truncating off the last coefficient in the case where $m_{i+1}
=2m_i-1$.

In a model where only roots of unity of order $2^k$ are allowed, it is
possible to use van der Hoeven's Truncated Fourier
Transform~\cite{Hoeven04}. For $f\in\K[x]$ of degree less than $m$,
let $\TFT(f,m)$ denote the values
$(f(w^{[0]_r}),\dots,f(\omega^{[i_{m-1}]_r}))$, where $r=\lceil
\log_2(m)\rceil$, $\omega$ is a primitive root of unity of order
$2^r$, and $[i]_r$ is the bitwise mirror of $i$ in length $r$.

A first difficulty is that the relationship between $\TFT(f,m)$ and
$\TFT(f,2m)$ is less transparent than in the case of the classical
Fourier transform. Step~2.a$'$ requires to compute only the values
$\TFT(f,2m)-\TFT(f,m)$; while it is obviously possible to adapt van
der Hoeven's algorithm to this case, as in~\cite[\S~5]{Hoeven05},
determining the exact cost requires a specific study. A second issue
is that using the values $\TFT(f,m)$ does not allow immediately to
perform multiplication modulo $x^m-1$, which is needed to compute $s$
at Step~2.d$'$ of our algorithm. However, this problem can be solved
by computing $s/x^m$, which is a polynomial of degree less than $m$
(remark that the same issue arises if one wants to use the Truncated
Fourier Transform in the algorithm of~\cite{HaZi04}).

\begin{figure}[!!!!!!!!!!!!!!!h]
\begin{center}
	\ifpdf
	\includegraphics[width=7cm]{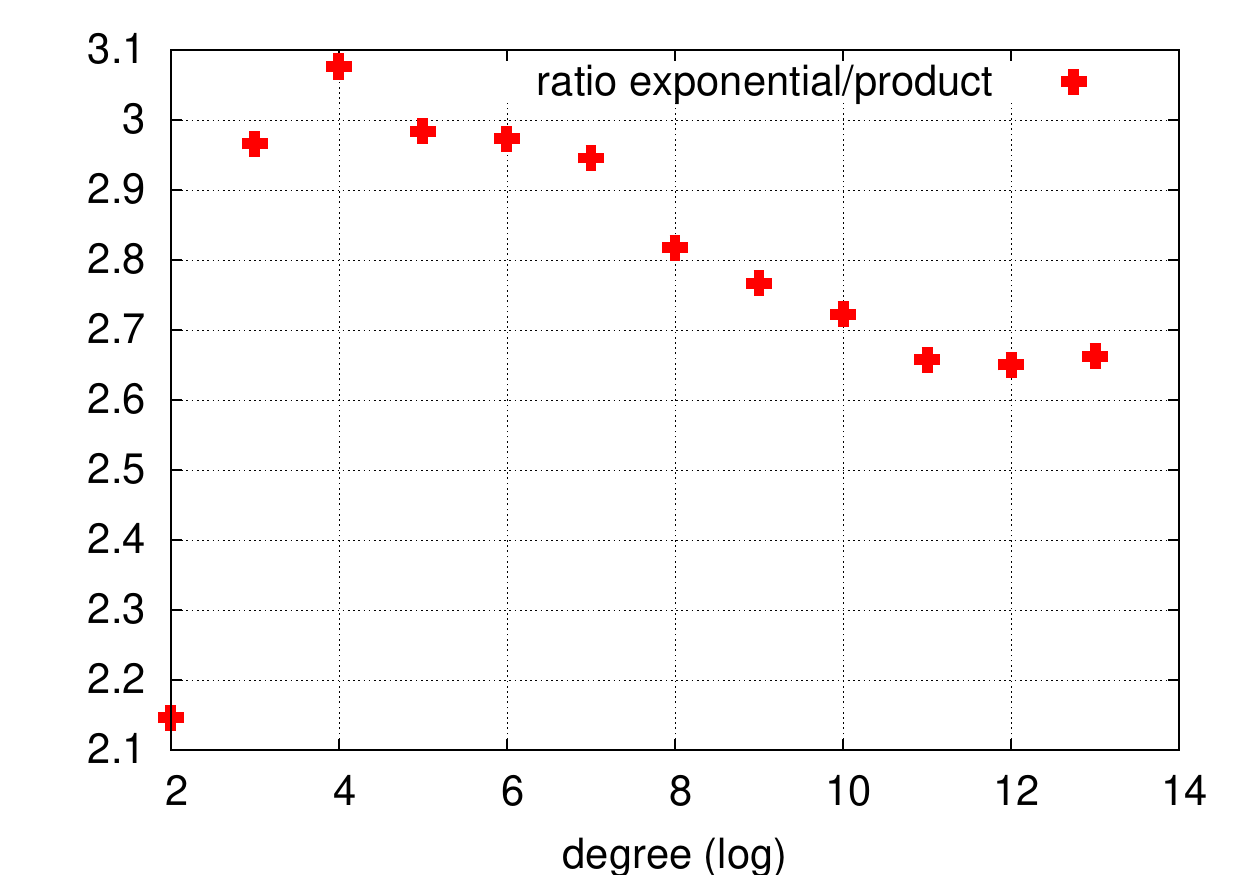}	
	\else
	\includegraphics[width=7cm]{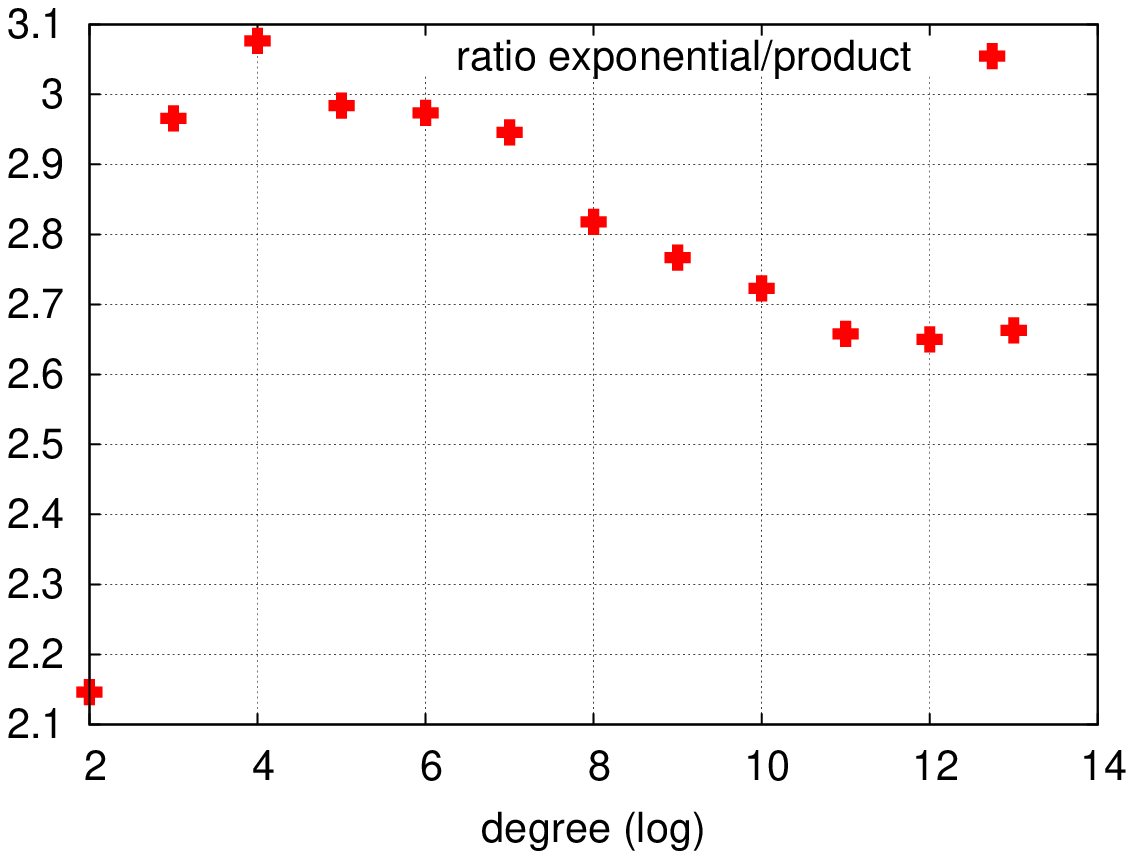}
	\fi
\end{center}	
\caption{Ratio exponential vs. product}
\label{fig:1}
\end{figure}

\paragraph*{Experiments.}
Figure~\ref{fig:1} gives empirical results, using the FFT routines
for small Fourier primes implemented in Shoup's NTL library~\cite{NTL}.
As can be seen, a ratio close to the expected 2.75 is observed.

\paragraph*{Acknowledgments.}   
We thank an anonymous referee for several useful remarks. This work was
supported in part by the French National Agency for Research (ANR Project 
``Gecko"), the joint Inria-Microsoft Research Centre, NSERC and the Canada 
Research Chairs program.        

\bibliographystyle{plain}

\end{document}